\documentclass[oldversion]{aa}  
\usepackage{graphicx}
\usepackage{txfonts}

\newcommand{\rsub}{R_{\rm sub}}
\newcommand{\Tsub}{T_{\rm sub}}
\newcommand{\Luv}{L_{\rm UV}}
\newcommand{\fAD}{f_{\rm AD}}
\newcommand{\tauk}{\tau_{\rm K}}
\newcommand{\Rtauk}{R_{\tauk}}
\newcommand{\Rin}{R_{\rm in}}
\newcommand{\jh}{$J$$-$$H$}
\newcommand{\hk}{$H$$-$$K$}

\begin{document}

\title{The innermost region of AGN tori: implications from the
  HST/NICMOS Type 1 point sources and near-IR reverberation}

\subtitle{}

\author{Makoto Kishimoto\inst{1},
       Sebastian F. H\"onig\inst{1},
       Thomas Beckert\inst{1}
       \and
       Gerd Weigelt\inst{1}
       }

\offprints{M. Kishimoto}

\institute{
           $^1$Max-Planck-Institut f\"ur Radioastronomie, Auf dem H\"ugel 69,
           53121 Bonn, Germany\\
           \email{mk@mpifr-bonn.mpg.de}
          }

\date{accepted by Astronomy \& Astrophysics on August 29, 2007}

\authorrunning{Kishimoto et al.}

\titlerunning{The innermost region of AGN tori}

 
\abstract { Spatially resolving the innermost region of the putative
torus-like structure in an active galactic nucleus (AGN) is one of the
main goals of its high-spatial-resolution studies.  This could be done
in the near-IR observations of Type 1 AGNs where we see directly the
hottest dust grains in the torus.  We discuss two critical issues in
such studies.  One is the possible contribution from the central
putative accretion disk (the near-IR part of the big blue bump
emission), which should be taken into account for the torus
measurements.  The other is the expected size of the inner boundary of
the torus, essential for the feasibility of spatially resolving the
region.

Firstly, we examine the nuclear near-IR point sources in the
HST/NICMOS images of nearby Type 1 AGNs, to evaluate the accretion
disk contribution.  After the subtraction of the host bulge flux
through two-dimensional decompositions, we show that near-IR colors of
the point sources appear quite interpretable simply as a composite of
a black-body-like spectrum and a relatively blue distinct component as
expected for a torus and an accretion disk in the near-IR,
respectively.  The near-IR colors of our radiative transfer models for
clumpy tori also support this simple two-component interpretation.
The observed near-IR colors for the available sample suggest a
fractional accretion disk contribution of $\sim$25\% or less at 2.2
$\mu$m.

Secondly, we show that the innermost torus radii as indicated by the
recent near-IR reverberation measurements are systematically smaller
by a factor of $\sim$3 than the predicted dust sublimation radius with
a reasonable assumption for graphite grains of sublimation temperature
1500~K and size 0.05~$\mu$m in radius.  The discrepancy might indicate
a much higher sublimation temperature or a typical grain size being
much larger in the innermost tori, though the former case appears to be
disfavored by the observed colors of the HST point sources studied
above.  Alternatively, the central engine radiation might be
significantly anisotropic.  The near-IR interferometry with a baseline
of $\sim$100 m should be able to provide the important, independent
size measurements for the innermost torus region, based on the low
fractional contribution from the accretion disk obtained above.

}


\maketitle
%

\section{Introduction}

The emission from active galactic nuclei (AGN) in the near- and
mid-infrared (IR) wavelengths is generally thought to be dominated by
the thermal emission from dust grains in the putative torus-like
structure surrounding the broad emission line region and the central
engine.  One of the main goals of high spatial resolution AGN studies
is to resolve the innermost region of these tori.  The innermost
structure in itself is physically interesting, but also the region
could well be closely related to and involved in the feeding process
of the central engine.  Spatially resolving the innermost torus could
be done in Type 1 AGNs where our line of sight is more or less close
to the symmetry axis direction of the torus and so the innermost
region is directly seen.  While IR interferometric observations have
resolved at least some part of this torus structure in nearby AGNs
(Wittkowski et al. 1998, Weinberger et al. 1999, Swain et al. 2003,
Weigelt et al. 2004, and Wittkowski et al. 2004 in the near-IR; Jaffe
et al. 2004 and Tristram et al. 2007 in the mid-IR; see below for more
discussions), most of the results obtained so far are for Type 2
objects where the symmetry axis of the torus is thought to have large
inclinations to our line of sight and so the innermost region of the
torus is not directly seen.

The inner boundary of the torus is thought to be set by dust
sublimation, and since the dust sublimation temperature is thought to
be roughly $\sim$1500K, the innermost region is mainly emitting in the
near-IR.  However, in Type 1s, some sizable part of the near-IR
emission comes from the central engine, which is also directly seen
along our line of sight, and this will affect the measurement of the
innermost region of the torus.  This fractional contribution is one of
the key quantities in high spatial resolution studies of the
innermost tori.  In this paper, we try to quantify the contribution
and its effect on the interferometric observations, based on the high
spatial resolution HST/NICMOS images of nearby type 1 AGNs which
minimizes the effect of host galaxies.

Another key quantity is the expected innermost size of the torus.  A
theoretical prediction of dust sublimation radii $\rsub$ by Barvainis
(1987) is quite a robust estimation of the inner torus boundary size.
On the other hand, this innermost radii can observationally be probed
by time-lag measurements between the UV/optical and near-IR. The
time-lag radii have recently been shown to be proportional to the
square root of the optical luminosity (Suganuma et al. 2006), which is
consistent with the prediction for dust sublimation radii. Therefore,
a direct comparison between these two radii should be very important.
We aim to discuss the implications from the comparison, and evaluate
the current expectation for near-IR interferometry based on the
expected innermost torus size.

In section 2, we will describe the available HST/NICMOS sample and the
measurement of the point source fluxes at multiple wavebands through
two-dimensional decomposition.  In section 3, these point-source
fluxes will be compared with simple blackbody colors and also with the
colors of more realistic clumpy torus models, to evaluate the
accretion disk contribution.  In section 4, we examine the expected
innermost size of the torus.  Based on these discussions, we aim to
clarify the current expectation for the interferometry of the
innermost torus region in section 5. We summarize our discussions in
section 6. When needed, cosmological parameters are assumed as
$H_0=70$ km s$^{-1}$ Mpc$^{-1}$, $\Omega_{\rm m}=0.3$, and
$\Omega_{\Lambda}=0.7$ throughout this paper.

   \begin{table*}
      \caption[]{The list of $z < 0.2$ radio-quiet Type 1 AGNs with
      simultaneous $J/H/K$ or $H/K$ observations in the HST/NICMOS
      archive. Also given are the physical scale for 0.21 arcsec (FWHM
      of the PSF in F222M), the measured flux for the central PSF
      component, and its fraction within a 2 arcsec diameter
      aperture.}
         \label{tab-flux}
     $$ 
         \begin{tabular}{lccccccc}
            \hline
            object name  & z & scale (pc)    & obs date & chip/filter & PSF flux & fraction (\%) \\ 
                         &   & for $0.''21$$^{\mathrm{a}}$ &   &      & (mJy)    & in 2 arcsec   \\ 
            \hline
            NGC 4151  & 0.00332 &  19 & 1998-05-22 & NIC2/F110W & 59.9 & 77 \\ 
                      &         &     &            & NIC2/F160W & 100. & 82 \\ 
                      &         &     &            & NIC2/F222M & 197. & 87 \\ 
            NGC 3227  & 0.00386 &  22 & 1998-04-06 & NIC2/F160W & 7.84 & 20 \\ 
                      &         &     &            & NIC2/F222M & 16.6 & 35 \\ 
            NGC 7469  & 0.0163  &  67 & 1997-11-10 & NIC2/F110W & 17.1 & 64 \\ 
                      &         &     &            & NIC2/F160W & 38.0 & 68 \\ 
                      &         &     &            & NIC2/F222M & 79.8 & 80 \\ 
            IC4329A   & 0.0161  &  75 & 1998-05-21 & NIC2/F160W & 47.5 & 75 \\ 
                      &         &     &            & NIC2/F222M & 107. & 85 \\ 
            NGC 5548  & 0.0172  &  79 & 1998-02-15 & NIC2/F160W & 16.5 & 68 \\ 
                      &         &     &            & NIC2/F222M & 32.7 & 79 \\ 
            Mrk 231   & 0.0422  & 182 & 1998-09-25 & NIC1/F110M & 24.9 & 81 \\ 
                      &         &     &            & NIC2/F160W & 68.6 & 84 \\ 
                      &         &     &            & NIC2/F207M & 130. & 88 \\ 
            IRAS07598+6508& 0.148& 563& 1997-11-11 & NIC2/F110W & 7.31 & 88 \\ 
                          &      &    &            & NIC2/F160W & 17.3 & 90 \\ 
                          &      &    &            & NIC2/F222M & 44.3 & 91 \\ 
            Mrk 1014  & 0.163    & 603& 1997-12-13 & NIC2/F110W & 1.27 & 53 \\ 
                      &          &    &            & NIC2/F160W & 3.01 & 61 \\ 
                      &          &    &            & NIC2/F222M & 6.93 & 68 \\ 

            \hline
         \end{tabular}
     $$ 
     \begin{list}{}{}
     \item[$^{\mathrm{a}}$] Calcualated from the radial velocities 
      corrected with a CMB dipole model from NED.
     \end{list}
   \end{table*}

     \begin{figure} \centering
     \includegraphics[width=9cm]{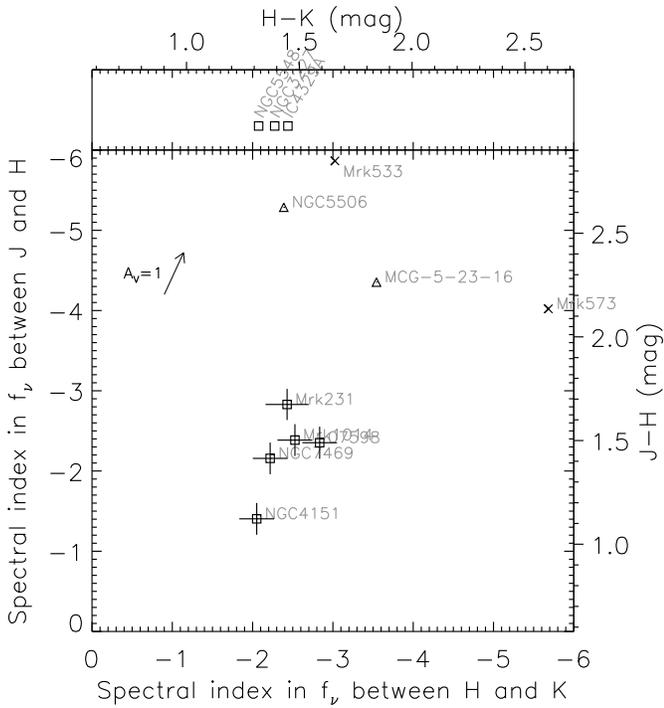} 

     \caption{The observed \jh\ and \hk\ colors for the nuclear point
     source in the HST/NICMOS images of the Type 1 AGNs listed in
     Table~\ref{tab-flux} are shown as squares with error bars. The
     colors are presented as the spectral index $\alpha$ in $f_{\nu}$
     ($\propto \nu^{\alpha}$) from the fluxes measured with F110W and
     F160W filters for \jh\ color, and F160W and F222M filters for
     \hk\ color. The corresponding $J$-$H$ and $H$-$K$ colors in Vega
     magnitude are shown as second axes at top and right (adopted zero
     points are 1848, 1084, 635 Jy for F110W, F160W, F222M,
     respectively). In the top small panel, the objects with only an
     \hk\ color are shown.  The colors of Type 2 (crosses) or
     intermediate-type objects (triangles) from Alonso-Hererro et
     al. (2001) have also been plotted for comparison with the Type 1
     objects. A foreground reddening vector for $A_V=1$ is indicated
     in the upper-left.}

     \label{jh-hk-obs}
     \end{figure}

\section{The HST/NICMOS images and point source flux}

\subsection{The sample and data reduction}

We have searched the HST archive for NICMOS observations of Type 1
AGNs which are in the AGN catalog compiled by V{\'e}ron-Cetty \&
V{\'e}ron (2006) with spectral classifications S1.0$-$S1.5 and
redshift $z$ up to 0.2, and have simultaneous observations at $J/H/K$
or $H/K$ bands. Filters are mostly F110W, F160W, and F222M, with a
central wavelength of 1.12, 1.60 and 2.22 $\mu$m, respectively.  We
further excluded radio-loud objects to avoid any possible contribution
in the near-IR from synchrotron components.  Table~\ref{tab-flux}
lists the objects found. For this sample, the physical linear size
corresponding to the FWHM of the NICMOS Point Spread Function (PSF) at
2.2 $\mu$m is less than $\sim$600 pc (see Table~\ref{tab-flux}). The
data were reduced and calibrated using the standard pipeline software
{\sc CALNICA} and {\sc CALNICB} with the most recent reference files
as of early 2007.  The detector readout mode was MULTIACCUM. In this
mode, each pixel in NICMOS detectors is read multiple times
non-destructively during a single integration.  The final count rate
at each pixel is calculated in {\sc CALNICA} by a least squares fit to
the accumulating counts versus exposure times at each readout. This
uses only non-saturated periods during the course of the single
exposure for each pixel.  Therefore, the brightest nuclear pixels have
effectively short exposure times and do not suffer from saturation.

\subsection{Nuclear flux measurements}

We implemented two-dimensional decompositions for the central $\sim$ 4
$\times$ 4 arcsec$^2$ of each image using our software written in
IDL. We used a model PSF generated by TinyTim (see Krist et al. 1998)
for the nuclear point source and a model bulge component for an
underlying host galaxy, convolved with the same model PSF, assuming de
Vaucouleurs $r^{1/4}$ profile. We then measured the total flux of the
PSF component.  

Generally, the point sources in NICMOS images can be well modeled by
the TinyTim synthetic PSFs, but detailed PSF shapes depend on several
factors, which have been investigated substantially by Krist et
al. (1998) and Suchkov \& Krist (1998).  To roughly estimate the
discrepancies between synthetic and observed PSFs, we compared
synthetic PSFs with the NICMOS images of the Galactic stars SA 107-626
and SA 107-627 taken with NIC1/NIC2 cameras with a few filters. These
have been observed for the PSF measurements by Kukula et al. (2001;
see also Veilleux et al. 2006 for notes on the latter star).  For each
stellar observation, we compared radial profiles of three images: (A)
an absolute difference image between the synthetic and observed PSF
with matched total counts; (B) a 'standard-deviation' image
$\sigma_{\rm pix}$ for the observed PSF, where each pixel consists of
the standard deviation of the 3 $\times$ 3 pixel region centered on
that pixel; (C) a statistical error estimate image $\sigma_{\rm stat}$
for the observed PSF from the pipeline.  In the outer region, the
pixel values of the $\sigma_{\rm pix}$ image B approaches those of the
$\sigma_{\rm stat}$ image C as expected. In the inner region close to
the PSF center, $\sigma_{\rm pix}$ is much larger than $\sigma_{\rm
stat}$, since the former rather represents the extent of small spatial
scale structures of the PSF on top of the statistical photon noise.
We found that the overall profile of the difference image A is roughly
reproduced by taking the geometric mean $\sigma_{\rm mean}$ of image B
and C, i.e. $\sigma_{\rm mean} = \sqrt{\sigma_{\rm pix} \sigma_{\rm
stat}}$.  Thus the average deviation of the observed image from the
model PSF can approximately be described by this $\sigma_{\rm mean}$.

This would also be applicable to our AGN images since a relatively
smooth, underlying host component would not contribute much to the
$\sigma_{\rm pix}$ image except for its photon noise.  Therefore we
adopted this $\sigma_{\rm mean}$ image as an estimate for the average
deviation of each observed AGN image from the model. In the
two-dimensional fit, we weighted each pixel by $1/\sigma_{\rm
mean}^2$, after masking out obvious distinct structures in the host
galaxies such as starburst knots in a few cases (e.g. NGC 7469). The
$\sigma_{\rm mean}$ image can easily be generated from the observed
image and the $\sigma_{\rm stat}$ image from the pipeline software.
The radial profile of the $\chi^2$ image of the resulting fits turned
out to be rather flat, with $\chi^2$ values being roughly of order
unity in most cases, indicating adequate weighting and fit.  The
reduced $\chi^2$ of the fits are thus quite close to unity in most
cases.

To estimate the errors in our PSF flux measurements, we compared the
PSF flux recovered by the fits for the Galactic stars above with those
from a synthetic aperture photometry with aperture corrections.  We
found that the former reproduces the latter within $\sim$5\%. In the
fits for the AGN images of our sample, we also measured the residual
flux within a 2 arcsec diameter aperture after the subtraction of the
fitted PSF and host components, and found it on average to be less
than $\sim$5\% of the measured PSF flux.  Therefore we assume 5\%
uncertainty in our PSF flux measurements. This has been found to be
typically several times larger than the formal statistical error given
by the fit.

The results of the flux measurements are summarized in
Table~\ref{tab-flux}. The measured fraction of the PSF component
within a 2 arcsec diameter aperture for each object is also tabulated.
Similar measurements have been made by Alonso-Herrero et al. (2001)
for 5 objects in the list, though only two of them have simultaneous
$J/H/K$ observations, for which we have confirmed approximate
consistency of the results.  Scoville et al. (2000) studied the same
NICMOS colors of ultra luminous infrared galaxies including a few
objects in our sample, by measuring the nuclear fluxes within a 1.1
arcsec diameter aperture with the adjacent background subtracted. Our
two-dimensional decomposition measurements for the common objects are
roughly in agreement with their measurements.

The observed \jh\ and \hk\ colors for these Type 1 objects are plotted
in Fig.\ref{jh-hk-obs} in squares after a small correction for the
Galactic extinction using $E_{B-V}$ from NED. The objects only with a
\hk\ color are plotted in the small top panel.  For comparison, we
have also plotted the same colors of the nuclear point sources in
several nearby Type 2 or intermediate-type objects (crosses and
triangles, respectively; the latters are designated as S1i in the
catalog by V{\'e}ron-Cetty \& V{\'e}ron, meaning that broad lines are
detected in the IR). These are at $z<0.03$ and the fluxes have been
deduced by Alonso-Herrero et al. (2001) using both ground-based and
HST images. The colors of the Type 1 objects are concentrated in a
relatively narrow range, and generally much bluer than those of the
Type 2 and intermediate-type objects.

\begin{figure}[tbh]
     \includegraphics[width=9cm]{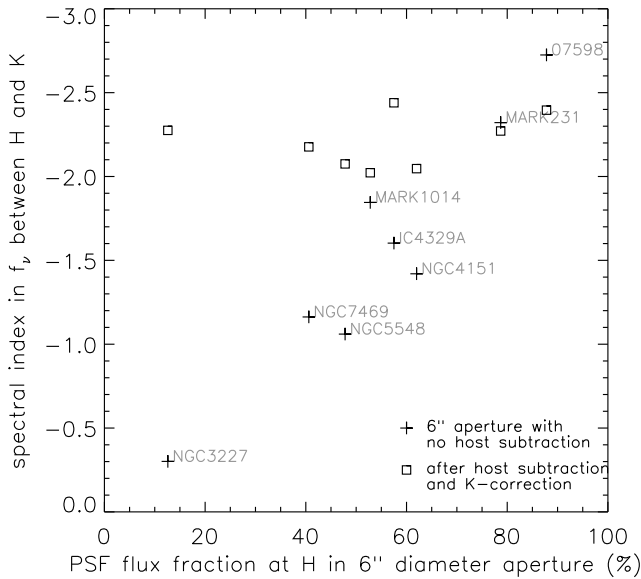} 

     \caption{\hk\ color versus PSF flux fraction in F160W filter
     image, with synthetic aperture of 6 arcsec diameter (plus
     signs). Plotted in squares are the \hk\ colors of the PSF-only
     flux, i.e. after host galaxy light subtraction (and also after a
     K-correction; see text).}

     \label{nucfrac}
     \end{figure}

To show the effect of the host subtraction (through the
two-dimensional PSF+host decomposition) on the color measurements as
compared with large-aperture measurements, we plotted the \hk\ color
versus the PSF flux fraction at $H$ band with a synthetic aperture of
6 arcsec diameter in Fig.\ref{nucfrac}. The color changes
significantly as a function of the PSF fraction.  For comparison, we
also plotted the \hk\ color of the PSF flux from the PSF+host
decomposition (the colors are the same as the \hk\ colors plotted in
Fig.\ref{jh-hk-kcor}, i.e. with a K-correction for objects with J/H/K
measurements; see section~\ref{sec-two-comp} below). As clearly seen,
the effect of the host subtraction using these high resolution images
is quite significant in the majority of the objects.

\section{The spectral components of the near-IR point sources}

\subsection{Hot dust emission and 
contribution from the putative accretion disk}\label{sec-two-comp}

Type 1 objects are thought to have low inclination angles, providing a
direct view of the central engine and the broad line region.  If we
assume that the stellar light from the host galaxy is well subtracted
by the PSF+host decomposition, the spectrum of the unresolved source
in the near-IR wavelengths is expected to be composed of essentially
two components: (i) the thermal emission from hot dust grains in the
innermost torus, nearly at the sublimation temperature; (ii) the
central engine emission, i.e. the near-IR tail of the so-called big
blue bump emission from the putative accretion disk. (We will discuss
other possible components below.)

\begin{figure*}[tbh]
     \centering 
       \includegraphics[width=14cm]{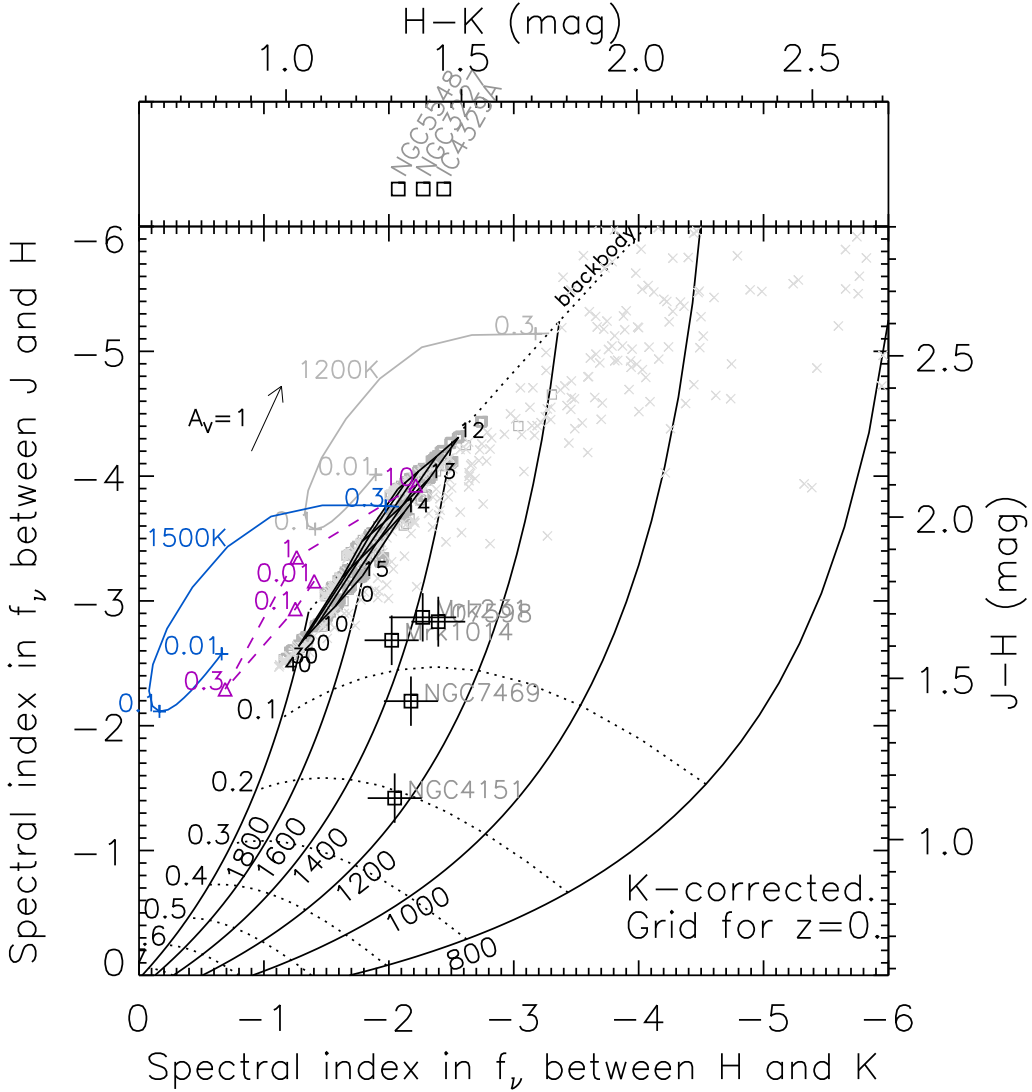}

       \caption{The same \jh\ and \hk\ colors for the Type 1 objects
       in Fig.\ref{jh-hk-obs} are shown but with a K-correction (see
       text; no-correction for the objects with only \hk\ color in the
       upper panel). A grid for \jh\ and \hk\ colors of a blackbody
       with temperature $T$ of 1800, 1600, 1400, 1200, 1000, and 800K,
       plus a blue power-law component with $f_{\nu} \propto
       \nu^{+1/3}$ which has a fractional contribution $\fAD$ at
       K-band from 0 with a 0.1 step, has also been plotted.  The \jh\
       and \hk\ spectral indices for each set of $(T, \fAD)$ are
       calculated using the transmission curves of F110W, F160W and
       F222M filters for the NIC2 camera. The small gray squares
       concentrated near the blackbody point of $T$$\sim$1400-1800K
       are the colors of our clumpy torus models with Type 1
       inclinations, while the small gray crosses which are spread
       toward much redder colors are for intermediate or Type 2
       inclinations.  The small flat grid overlayed on the small gray
       squares, labeled with inclinations (0$-$40\degr) and
       temperature (12$-$15, in units of 100K) is the average colors
       for Type 1 models. The tracks of colors for optically-thin
       graphite grains with different sizes are plotted in gray/blue
       curves for $T$=1200 and 1500K, where the colors for grain radii
       $a =$ 0.01, 0.1, 0.3 $\mu$m are marked as plus signs. The same
       colors for silicate grains with $T$=1500K and $a =$ 0.01, 0.1,
       0.3, 1, and 10 $\mu$m are marked as triangles and connected
       with gray/purple dashed lines.}

       \label{jh-hk-kcor} 
     \end{figure*}

If we simply assume the near-IR spectrum of dust grains to be of a
blackbody with a single temperature $T$, and that of an accretion disk
to be of a power-law form $f_{\nu} \propto \nu^{+1/3}$ which is a long
wavelength limit of simple multi-temperature blackbody disks (Shakura
\& Sunyaev 1973), then the spectral shape in the rest frame is fixed
for a given $T$ and a given fraction of the accretion disk
contribution at a certain wavelength.  We parameterize the latter as
the disk fraction at the rest wavelength of 2.2 $\mu$m and denote it
as $\fAD$. These $T$ and $\fAD$ can be calculated from the observed
set of \jh\ and \hk\ colors and redshift $z$ for each object.
Equivalently, we can produce a grid of $T$ and $\fAD$ on the \jh\ and
\hk\ color plane for $z=0$, and plot the \jh\ and \hk\ colors of each
object with K-corrections assuming this two-component spectrum. Such a
color-color diagram is shown in Fig.\ref{jh-hk-kcor} for the Type 1
objects. Note that such K-corrections are noticeable essentially only
for two objects in the plot.  Fig.\ref{jh-hk-kcor} shows that the
observed colors follow a relatively well-defined trend: they are along
a locus of a roughly similar temperature 1200$\sim$1500K for the hot
dust component, with a small range in $\fAD$ of 5$\sim$25\%. The
corresponding accretion disk fractions at $H$ and $J$ are much higher,
0.1$\sim$0.5 and 0.4$\sim$0.9, respectively.

\subsection{The near-IR colors of clumpy torus models}\label{sec-model}

To estimate the extent of the difference from simple black-body
spectra, we have simulated the near-IR colors of more realistic AGN
tori using our clumpy torus model (H\"onig et al. 2006; Beckert \&
Duschl 2004). We concentrate on Type 1 cases where inclinations $i$
are smaller than the half opening angle of the torus and thus our line
of sight toward the central engine is free of any model cloud.

We calculated spectra with a dust sublimation temperature $\Tsub$ of
1200$-$1500K and a radial distribution of the number of clouds
$\eta_r(r) \propto r^{\beta}$ where $\beta=-1.1 \sim -2.0$. For each
set of model parameters, we calculated 10 random arrangements of cloud
placements with a fixed opening angle, or more precisely, a fixed
ratio of scale height to radius corresponding to an average half
opening angle of $\sim$40\degr\ (see H\"onig et al. 2006 for more
details). We measured colors for the inclination angles free of clouds
along the line of sight. These colors are plotted in small gray
squares in Fig.\ref{jh-hk-kcor}.  Overlayed on these gray square
points is a small grid of $(T,i)=(1200$$-$$1500$K, $0$$-$$40$\degr$)$
with $\beta$=$-1.5$, which shows the averaged colors of the 10 random
arrangements for each set.  The comparison of this single-radial-index
grid with the gray square points (which include the indices from
$-1.1$ to $-2.0$) shows that the radial-index generally has a smaller
effect on colors for this half-opening angle case with Type 1
inclinations.  As the inclination increases from 0\degr\ to 40\degr,
the colors become slightly bluer since the fractional contribution
from the hotter side of the innermost clouds becomes more
significant. This trend holds until the inclination becomes close to
the half opening angle of the torus, when the colors start to become
much redder.  For comparison we also have plotted the model colors of
Type 2 or intermediate-type inclinations ($i \ga 40$\degr) for which
one or more clouds are along the line of sight, as small gray
crosses. They generally show much redder colors and much larger
scatters than those of Type 1 cases.

We also have calculated the colors of optically-thin emission from
dust grains with various sizes (radii $a$ from 0.001 to 10 $\mu$m) for
two temperatures, 1200 and 1500K, using absorption efficiencies
calculated by Draine \& Lee (1984) and Laor \& Draine (1993).  The
color tracks for graphite grains are plotted in Fig.\ref{jh-hk-kcor}
in gray/blue curves, where colors for $a$=0.01, 0.1, 0.3 $\mu$m are
marked as plus signs. The colors of silicate grains are shown only for
$T$=1500K and $a$=0.01, 0.1, 0.3, 1, and 10 $\mu$m cases in triangles
for clarity, and they are connected with gray/purple dashed lines.  As
expected, the colors approach the blackbody colors of the same
temperature as the grain size becomes larger.  The curves for graphite
grains are truncated at $a$=0.35 $\mu$m, beyond which the colors are
almost the same as that of a blackbody.

The spectra of our clumpy tori essentially corresponds to the case of
$a$=0.05 $\mu$m for $\sim$1:1 mixture of graphite and silicate grains
(H\"onig et al. 2006).  Then the near-IR colors of the clumpy tori
with a given $\Tsub$ can be seen as sitting roughly between the
blackbody and the optically-thin case (close to graphite grains with
$a$=0.05 $\mu$m, due to a much larger absorption efficiency of
graphites than that of silicates in near-IR) with $T=\Tsub$.  In more
detail, the overall \jh\ colors of the clumpy tori with a given
$\Tsub$ are roughly the same as the optically-thin color with
$T=\Tsub$, because the \jh\ wavelength region is dominated by the
emission from the hottest grains.  On the other hand, the \hk\ colors
are slightly redder, somewhere between the optically-thin and
blackbody case, since it has some contribution from slightly cooler
grains at $K$-band.  Because of this relatively simple behavior, our
results can easily be extrapolated toward slightly higher and lower
$\Tsub$ cases, and the results for a particular grain size can also be
roughly inferred.  If, for example, the grains in the innermost torus
is dominated by those with $a$ much larger than 0.05 $\mu$m, which we
actually argue in the next section, the torus near-IR color grid shown
in Fig.\ref{jh-hk-kcor} will slightly shift to a redder side (both in
\jh\ and \hk), much closer to the corresponding blackbody colors.

In all cases, the overall colors of Type 1 torus models stay quite
close to the locus of blackbody colors.  Therefore, the above estimate
of the fraction of the accretion disk contribution, $\fAD \la 25$\%,
essentially stays the same.  

The observed colors of the HST point sources constrain the plausible
range of $\Tsub$ at least to some extent. The colors appear to
disfavor the cases with $\Tsub$ much higher than $\sim$1500K. This
would be true even in the large-grain case described above.

\subsection{The near-IR spectral shape of the big blue bump} 

In the estimation above, we have assumed the near-IR spectral shape of
the big blue bump (BBB) as $f_{\nu} \propto \nu^{+1/3}$.  The near-IR
part of the BBB is still quite unknown, simply because we usually
cannot measure it due to the strong dominance of the thermal dust
emission from the torus. However, there are a few pieces of
observational evidence that the near-IR part of the BBB is quite blue,
as blue as the long wavelength limit of a non-truncated standard
accretion disk, $f_{\nu} \propto \nu^{+1/3}$, which is much bluer than
the observed optical BBB shape (e.g. $f_{\nu} \propto \nu^{-0.2}$;
Neugebauer et al. 1987; Francis et al. 1991).  One piece of evidence
is from a near-IR polarization study of one quasar by Kishimoto et
al. (2005). They show that the near-IR polarized flux spectrum, which
is argued to represent the intrinsic shape of the near-IR BBB, is of
the form $f_{\nu} \propto \nu^{+0.42\pm0.29}$.  Another is from the
reverberation measurements between the optical and near-IR for one
nearby Seyfert 1 galaxy by Tomita et al. (2006). They show that the
putative disk component has a spectral form of $f_{\nu} \propto
\nu^{-0.1 \sim +0.4}$.  Of course, these are still far from measuring
the spectral shape for the whole population, but pointing towards a
color bluer than the observed optical BBB shape.  The detailed form of
the disk component essentially does not affect our conclusions, as
long as it remains relatively blue.  The disk component color is the
convergence point of the grid in Fig.\ref{jh-hk-kcor} (lower left,
outside of the figure), and if the color is bluer than e.g. the
observed optical BBB shape cited above, which is quite likely, an
estimate for $\fAD$ will not change significantly.

\subsection{Other components}

One potential contamination by an additional component in the near-IR
would be a synchrotron emission from radio jets, but this is quite
unlikely since we have restricted our objects to radio-quiet ones. On
the other hand, it is possible that the NICMOS PSF fluxes have some
contribution from a nuclear young stellar cluster on top of the host
galaxy profile extrapolated from larger scales. In fact, recent
high-spatial-resolution integral field spectroscopic studies (Davies
et al. 2007 and references therein) indicate the existence of such
young stars with estimated ages of 10-300 Myr in the nuclear vicinity
of nearby Seyfert galaxies, including a few objects in our sample (NGC
3227, NGC 7469, Mrk 231). However its effect on our estimation of
$\fAD$ is expected to be quite small based on the following
reasons. (1) Its luminosity contributes only a few \% in
$\sim$10-100pc scales in most of the Type 1 objects in their
studies. (2) The near-IR colors of these young stars are much bluer --
the spectral index for \jh\ and \hk\ is $\alpha_{JH}\sim-0.5$ and
$\alpha_{HK}\sim+0.5$, respectively (Scoville et al. 2000, Fig.5) --
than the observed colors of the NICMOS point sources studied here.
Even if these stars are significantly reddened by internal dust
grains, which could be the case for a few luminous infrared galaxies
in our sample (Mrk 231, IRAS07598+6508, Mrk 1014), the reddened color
($\alpha_{JH}\sim-2.0$, $\alpha_{HK}\sim-0.5$; Scoville et al. 2000)
is still bluer than the observed point-source colors at least in
$H$$-$$K$.  Therefore, the subtraction of these young stellar
component can only make the remaining components redder, which might
even decrease the estimation of $\fAD$, not increase.

A possible foreground reddening of the nuclear point source light can
lead to an underestimation of $\fAD$ (Fig.\ref{jh-hk-kcor}).  At least
an approximate upper limit on the foreground reddening can be obtained
from the broad line ratios.  No significant reddening in the ratios
($A_V<0.5$) has been obtained for NGC4151, NGC5548, NGC7469 (Lacy et
al 1982) and Mrk1014 (Wu et al. 1998). For the rest of the sample,
some reddening is present: $A_V \sim 1.5$ for Mrk 231 (Lacy et
al. 1982), $A_V \sim 1.4$ for IRAS07598+6508 (Hines \& Wills 1995),
and $A_V \sim 2-4$ for IC4329A (due to an edge-on dust lane, Winkler
et al. 1992 and references therein). The estimation of $\fAD$ for
these objects slightly increases accordingly, but is still consistent
with $\fAD \la 25$\%.

Therefore, we conclude that, at least for the sample studied here, the
fractional contribution from an accretion disk at K-band is less than
$\sim$25\%.  Similar conclusions have been obtained from a detailed
analysis of the near-IR reverberation data for a Seyfert 1 galaxy
(Tomita et al. 2006) and from detailed near-IR spectroscopy of quasars
(Kobayashi et al. 1993), suggesting the validity of our simple
approach.

     \begin{figure}
     \centering 
       \includegraphics[width=9cm]{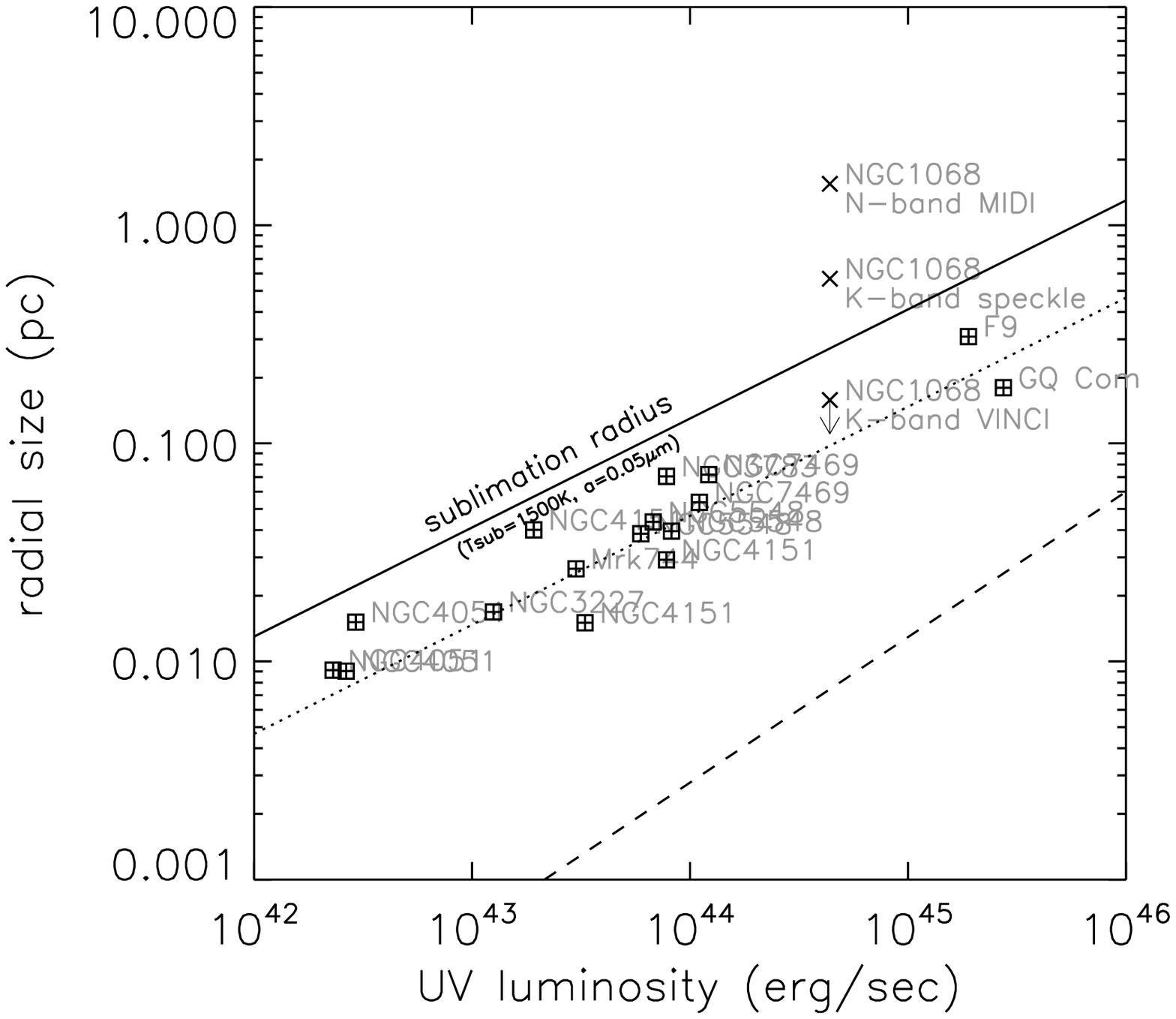}

       \caption{Dust sublimation radius given by Barvainis (1987;
       solid line) for $\Tsub$=1500K and grain size $a$=0.05 $\mu$m as
       a function of UV luminosity $\Luv$, as compared with K-band
       time-lag radii $\Rtauk$ for various Type 1 AGNs (Suganuma et
       al. 2006 and references therein). For the latter, $\Luv=6 \ \nu
       L_{\nu}(V)$ is assumed, and a fit in $\Rtauk \propto L^{1/2}$
       given by Suganuma et al. is shown in dotted line. The dashed
       line is a rough estimation of K-band emitting radii for an
       untruncated standard accretion disk. Also plotted are the
       interferometrically-measured radial sizes approximately
       perpendicular to the innermost linear radio structure for NGC
       1068. The UV luminosity for NGC 1068 here is an unobscured
       (thus estimated) value.}

       \label{torus_pc}
     \end{figure}

     \begin{figure}
     \centering 
       \includegraphics[width=9cm]{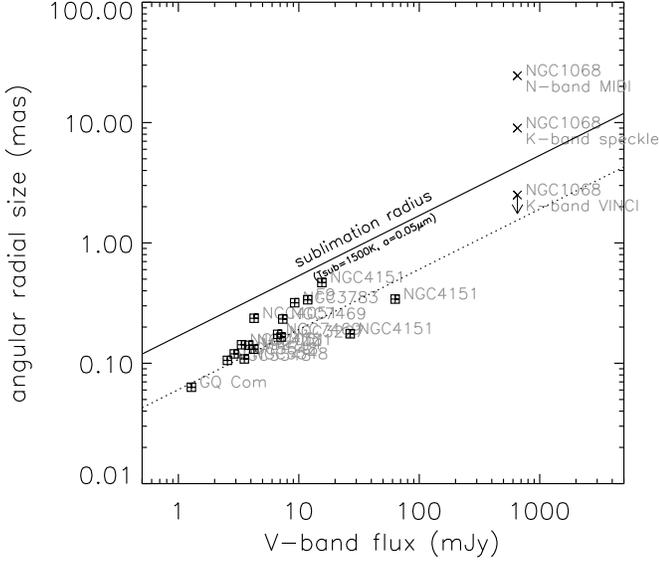}
       \caption{The same comparison as in Fig.\ref{torus_pc} but in
       angular radius versus V-band flux, with the solid and dotted
       line corresponding to equation~\ref{eq-Rsub-mas} and
       \ref{eq-Rtauk-mas}, respectively (i.e. $z \ll 1$ cases). The
       V-band flux for NGC 1068 here is an unobscured (thus estimated)
       value.}

       \label{torus_mas}
     \end{figure}

\section{The inner boundary of AGN tori}

In this section we focus on the expected physical and angular size of
the torus inner boundary, which is another critical issue in the high
spatial resolution studies of AGN tori.  We first summarize the dust
sublimation radius in physical and angular size, and compare it
with the time-lag radius from the near-IR reverberation measurements
in the literature, and then discuss the implications.

\subsection{Dust sublimation radius}

The inner boundary of tori is believed to be set by the sublimation of
dust grains. A theoretical calculation of dust sublimation radius
$\rsub$ by Barvainis (1987) is a well-referenced, quite robust
estimation of this inner boundary, which is given as
\begin{equation}
\rsub = 1.3 \ \left( \frac{\Luv}{10^{46} \ {\rm
    erg/sec}} \right)^{1/2}  
    \left( \frac{\Tsub}{\rm 1500K} \right)^{-2.8} 
    \left( \frac{a}{0.05 \mu m} \right)^{-1/2} \ {\rm pc}.
\end{equation}
In this calculation, Barvainis adopted absorption efficiencies for
graphite grains, since they have higher sublimation temperatures
$\Tsub$\ than silicate grains and thus are supposed to survive in
regions further in.  The value of $\Tsub$=1500K is also adopted by
Barvainis (1987).  For a plausible range of ambient gas pressures,
$\Tsub$\ for graphite and silicate grains are estimated to be
$\sim$1500$-$1900 K and $\sim$1000$-$1400 K, respectively
(e.g. Salpeter 1977, Huffman 1977, and references therein; Phinney
1989; Guhathakurta \& Draine 1989; Laor \& Draine 1993).  We
explicitly included the grain radial size $a$ of 0.05 $\mu$m assumed
by Barvainis (1987), which sits in the middle of the inferred size
distribution for interstellar graphite grains (Mathis, Rumpl \&
Nordsieck 1977).  The approximate proportionality $\rsub \propto
a^{-1/2}$ comes from the absorption efficiency $Q_{\rm abs}$ of a dust
grain being roughly proportional to its radius $a$ at a certain
wavelength in the near-IR (Draine \& Lee 1984).  $\Luv$ is the
UV-optical luminosity of the central engine, or more precisely, the
integration of the central engine luminosity weighted by $Q_{\rm abs}$
of a dust grain in question.

We can write a corresponding angular radius
$\theta_{\rsub}$ in milliarcsecond (mas) as
\begin{eqnarray}
\theta_{\rsub} 
& = & 1.2 \left( \frac{\Luv}{6 \ \nu L_{\nu}(V)}  \
\frac{f_{\nu}(V)}{50 {\rm mJy}}
\right)^{1/2} \nonumber \\
&& \ \ \ \ \cdot
 \left( \frac{\Tsub}{\rm 1500K} \right)^{-2.8} 
\left( \frac{a}{0.05 \mu m} \right)^{-1/2} 
\ {\rm mas}, 
\label{eq-Rsub-mas}
\end{eqnarray}
for $z \ll 1$ cases. Here $f_{\nu}(V)=50$mJy approximately corresponds
to the V-band flux of the nucleus of the brightest Seyfert 1 galaxy
NGC 4151.  $\Luv / 6 \nu L_{\nu}(V)$ is roughly unity for a generic
AGN spectral energy distribution (SED; Elvis et al. 1994; Sanders et
al. 1989). Note that the angular size is simply proportional to the
square root of observed flux for $z \ll 1$ cases.

\subsection{The near-IR reverberation radius}

The variability in the near-IR flux of several nearby Seyfert 1
galaxies has been observed to have a delay from the UV/optical
variability (Suganuma et al. 2006 and references therein), and the
time lags in these objects have been shown to be consistent with being
proportional to $L_{\rm opt}^{1/2}$ where $L_{\rm opt}$ is the optical
luminosity (Suganuma et al. 2006).  The light travel distance for the
time lag at K-band, $\Rtauk$, has been interpreted to be the distance
from the compact UV/optical source to the K-band emitting region,
which is thought to be the innermost region of the torus where dust
grains are nearly at their sublimation temperature.

The proportionality $\Rtauk \propto L_{\rm opt}^{1/2}$ is in a nice
agreement with the proportionality $\rsub \propto \Luv^{1/2}$ for a
generic AGN SED. However the values suggested by the time-lag radii
$\Rtauk$ and by $\rsub$ (with $\Tsub$=1500K and $a$=0.05 $\mu$m) are
actually quite different.  Fig.\ref{torus_pc} compares $\rsub$ and the
observed $\Rtauk$ in physical linear scale as a function of $\Luv$,
and Fig.\ref{torus_mas} in angular scale as a function of optical
V-band flux $f_{\nu}(V)$.  A fit to the time-lag data points given by
Suganuma et al. (2006) can be written as
\begin{equation}
\Rtauk = 0.47 \ \left( \frac{6 \ \nu L_{\nu}(V)}{10^{46} \ {\rm
erg/sec}} \right)^{1/2} \ {\rm pc},
\end{equation}
and the corresponding angular radius for $z \ll 1$ cases is given as
\begin{equation}
\theta_{\tauk} = 0.43 \ \left(  
\frac{f_{\nu}(V)}{50\ {\rm mJy}} \right)^{1/2} \ {\rm mas}.
\label{eq-Rtauk-mas}
\end{equation}
These fits are plotted in dotted lines in Fig.\ref{torus_pc}
and \ref{torus_mas}.  As can be seen clearly, the time-lag radii
are systematically {\it smaller} than $\rsub$ by a factor of
$\sim$3.

\subsection{Implications}

The case for the time-lag radius to be the radius of the dominant
K-band emitting region, which is thought to be the innermost torus,
seems likely, since the reverberation monitorings show that quite a
significant fraction of the K-band emission is varying in response to
the UV/optical in many Type 1 objects (e.g. Glass 2004; Minezaki et
al. 2004; Suganuma et al. 2006).  Some minor part of the factor 3
might be from a geometrically shortened delay, if the directly
illuminated region is only the surface or skin-like region of the
inner torus with a relatively small opening angle. However, this would
probably not be the dominant factor, as long as the most dominant
emission region is still roughly in the equatorial plane of the torus.

Then one implication might be that the sublimation temperature is much
higher than $1500$ K, and/or the typical grain radius is much larger
than 0.05 $\mu$m.  In fact, this has been inferred for some individual
time-lag cases (e.g. Clavel et al. 1989, Barvainis 1992 for Fairall 9;
Sitko et al. 1993 for GQ Com).  However, as we noted in
section~\ref{sec-model}, sublimation temperatures much higher than
$\sim$1500 K appear to be disfavored by the observed near-IR colors of
the HST point sources (Fig.\ref{jh-hk-kcor}). In this case, the above
comparison of $\rsub$ and $\Rtauk$ might simply suggest a dominance
of large grains in the innermost region, and this would systematically
be true (at least among these reverberation objects), not just in some
individual objects.  The sublimation radius as a function of $\Tsub$\
and grain size is well illustrated graphically in Fig.8 of Laor \&
Draine (1993). If the sublimation temperature is $\sim$1500 K, the
typical grain size in radius would be $\sim$0.2 $\mu$m to have $\rsub$
go down to match $\Rtauk$. This is quite large enough to be rather
close to the blackbody limit (i.e. a larger size will not change
$\rsub$ too much further), but still in the inferred size range for
Galactic interstellar graphite grains (e.g. Mathis et al. 1977).  Note
that a dominance of large grains in AGN tori has been suggested and
discussed on different grounds (e.g. Maiolino et al. 2001a,b; Gaskell
et al. 2004).

Another possibility is an intrinsic anisotropy of the central engine
radiation, with a factor of $\sim$10 reduction from a polar to
equatorial direction, i.e. much less luminosity toward the torus
(equatorial) direction than to our (polar) line of sight.  This might
be consistent with accretion disks where we expect the effects of
projected area and limb darkening (e.g. Netzer 1985). Alternatively
there might be a significant extinction, i.e. absorption and/or
scattering, of the central engine radiation before reaching the
torus. However, its consequences has to be carefully considered.  In
any case, the {\it systematic} difference between the time-lag radii
and $\rsub$ above indicates that dust properties and/or
anisotropy/extinction in the innermost region is somehow similar in
these various objects.

     \begin{figure}
     \centering \includegraphics[width=9cm]{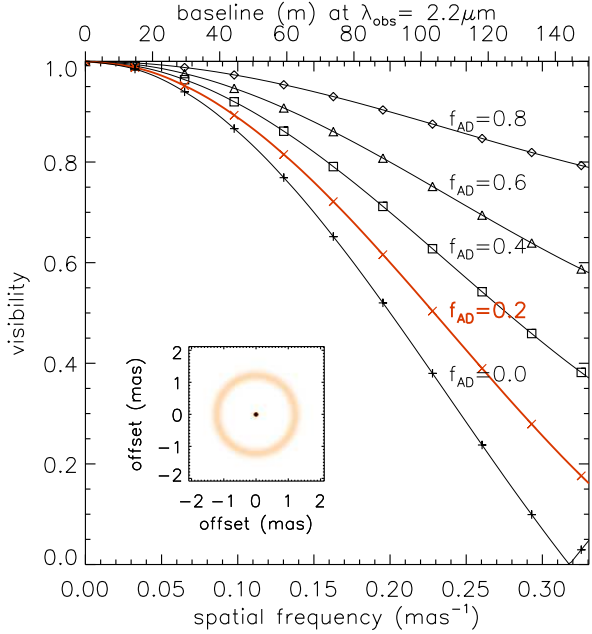}
       \caption{Visibilities, or normalized amplitudes of the Fourier
       transform, for a model surface brightness image of a 1.2 mas
       radius ring plus a much more compact source at the center with
       various flux fraction $\fAD$ of the latter. The baselines
       corresponding to the spatial frequencies in the bottom $x$-axis
       are shown in the top $x$-axis label for an observing wavelength
       of 2.2 $\mu$m. The plausible case of $\fAD$=0.2 is highlighted
       in red, and the corresponding model image is shown in the
       inset.}

       \label{vis_ring_1p2}
     \end{figure}

     \begin{figure}
     \centering 
       \includegraphics[width=9cm]{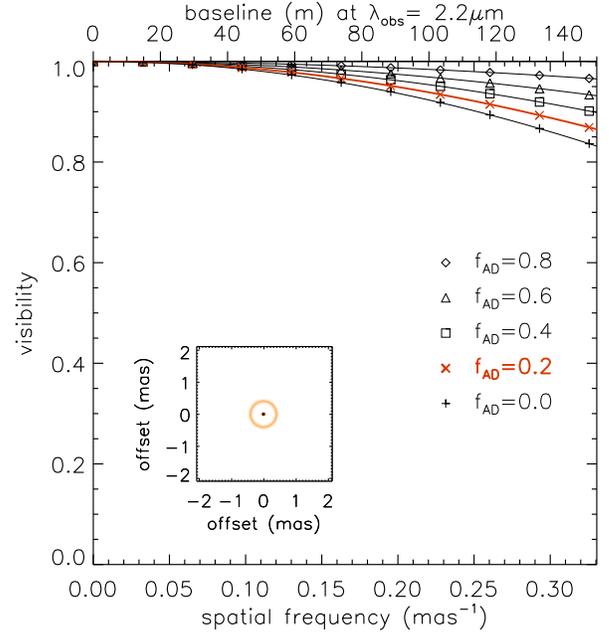}
       \caption{The same as Fig.\ref{vis_ring_1p2} but for a ring
       radius of 0.4 mas. The same symbols for each $\fAD$ are used.}

       \label{vis_ring_0p4}
     \end{figure}

\section{The near-IR interferometry of Type 1 nuclei} 

Small angular sizes of AGN tori require IR interferometric
observations to spatially resolve the structure, and some structures
actually have been resolved in a few objects with interferometry both
in the near-IR (Wittkowski et al. 1998, Weinberger et al. 1999,
Weigelt et al. 2004, Wittkowski et al. 2004 for NGC 1068; Swain et
al. 2003 for NGC 4151) and mid-IR (Jaffe et al. 2004 for NGC 1068;
Tristram et al. 2007 for Circinus).  Based on the discussion of the
accretion disk flux fraction and the angular size of the torus inner
boundary in the previous sections above, we discuss the current
expectations for the near-IR interferometric studies of the innermost
region of AGN tori, and also discuss the existing measurements.

\subsection{Simulated interferometric observations}

For the sample studied in sections 2 and 3, the fractional
contribution at K-band from a putative accretion disk is suggested to
be only $\sim$25\% or less, so that the point-source K-band flux is
essentially dominated by the flux from the torus. On the other hand,
the physical size of the near-IR emitting region in the accretion disk
is expected to be much smaller than the torus inner boundary.
Assuming a simple geometrically-thin optically-thick multi-temperature
blackbody disk (Shakura \& Sunyaev 1973), we can at least formally
estimate the K-band emitting radius $r_{\rm AD,K}$ as a radius where
the disk temperature goes down to $\sim$1500K. We obtain
\begin{equation}
r_{\rm AD,K} \simeq 0.06 \ \left( \frac{\eta}{0.1} \right)^{-1/3} 
\left( \frac{L}{10^{46} \ {\rm erg/sec}} \right)^{2/3}
\left( \frac{L/L_{\rm Edd}}{0.1} \right)^{-1/3} \ {\rm pc}, 
\end{equation}
where $\eta$ and $L_{\rm Edd}$ are the radiative efficiency (the
luminosity $L = \eta \dot{M} c^2$; $\dot{M}$ is the mass accretion
rate) and Eddington luminosity, respectively.  This is shown in
Fig.\ref{torus_pc} as a dashed line. The effective size of the K-band
emitting region might be even smaller due to the truncation of the
outer part of the disk by self-gravity (e.g. Goodman 2003).

Fig.\ref{vis_ring_1p2} and \ref{vis_ring_0p4} show the simulation of
interferometric observations for a simple case of a ring-like torus
plus a much more compact source, corresponding to an accretion disk,
with various fractional flux contributions $\fAD$.
Fig.\ref{vis_ring_1p2} is for a ring radius of 1.2 mas corresponding
to the angular size in Eq.~\ref{eq-Rsub-mas} ($\rsub$ case), and
Fig.\ref{vis_ring_0p4} is for 0.4 mas in Eq.~\ref{eq-Rtauk-mas}
($\Rtauk$ case).  Here, the FWHM $w$ of the ring is set to 1/5 of the
ring radius $r_{\rm ring}$, but the results for the spatial frequency
range shown in these Figures essentially do not change as long as $w
\ll r_{\rm ring}$.

The model image (shown in the insets), or the surface brightness
distribution, can apparently look dominated by the point source even
for an $\fAD \sim 0.2$ case, simply because of the small emitting size
of the point source.  However, in the Fourier-transformed image, the
low spatial frequency part of the Fourier amplitude profile, or
visibility profile, is essentially determined by the size of the torus
in the cases of low fractional contributions from the
disk. Fig.\ref{vis_ring_1p2} and \ref{vis_ring_0p4} show that, when
$\fAD$ goes down to $\sim$0.2, the profile is already quite similar to
the torus-only case (i.e. $\fAD$=0).  Simulations with different ring
radii show that, if $\fAD$ is confined to low values of $\la$0.2, the
dispersion of the visibility curves for a given ring radius caused by
this small range of $\fAD$ is roughly equivalent to the dispersion due
to a ring radius change by $\pm\sim$10\% or less.  Therefore, if
$\fAD$ is confined to low values of $\la$0.2, which seems to be the
case at least for the Type 1 objects studied here, the size
measurement of the torus in the K-band will not be significantly
affected by the accretion disk component.

For both cases with the inner boundary given by $\rsub$ (with
$\Tsub$=1500K and $a$=0.05 $\mu$m) and by $\Rtauk$,
Fig.\ref{vis_ring_1p2} and \ref{vis_ring_0p4} show that we can measure
the size of the innermost region of the torus with a baseline of
$\sim$ 100 m at K-band, which is achievable with existing facilities,
although we would need at least a few \% accuracy in the visibility
measurements for the latter case.  Note that the actual visibility for
a more realistic torus case would be smaller than the ring case whose
radius is equal to the inner boundary radius of the torus, since the
intensity distribution would be more extended outwards and not
inwards.

\subsection{Comparison with existing measurements}

For the nucleus of the Seyfert 1 galaxy NGC 4151, Swain et al. (2003)
measured a squared visibility $V^2$ of 0.84$\pm$0.06 (or
V=0.92$\pm$0.03) at a projected baseline of 83 m in K-band with the
Keck interferometer. They favored that most of the K-band emission is
coming from an unresolved accretion disk, rather than having a
ring-like torus emission as the dominant component. One major reason
was that, if a ring-like geometry is assumed, the ring radius implied
by the measured visibility is $\sim$0.5 mas ($\sim$0.04 pc) which is
much smaller than $\rsub$ for $\Tsub$=1500K and $a$=0.05 $\mu$m
($\sim$1.2 mas or 0.10 pc for NGC 4151; see
Fig.\ref{torus_mas}). However, this implied ring radius is essentially
consistent with the time-lag radius which is $\sim$0.4 mas for
NGC4151.  The Keck visibility measurement is consistent with the
visibility curves in Fig.\ref{vis_ring_0p4} -- note that $\fAD$ is
estimated to be $\sim$0.2$-$0.25 for NGC 4151 (Fig.\ref {jh-hk-kcor}).
Therefore the interpretation can drastically change if the time-lag
radius represents the actual innermost torus radius.

For the nucleus of the Seyfert 2 galaxy NGC 1068, a few
interferometric observation have been carried out. In the mid-IR, a
structure has been resolved with VLTI/MIDI (Jaffe et al. 2004) where a
Gaussian-modeled\footnotemark\ radial size perpendicular to the
innermost radio jet axis is $\sim$25 mas (HWHM; $\sim$1.6 pc). At
K-band, bispectrum speckle interferometry (Wittkowski et al. 1998;
Weigelt et al. 2004) has shown that the visibility goes down quite
quickly from unity to $\sim$0.7 at short baselines (0$\sim$6 m), with
a measured radial size perpendicular to jet being $\sim$9 mas
(Gaussian HWHM; $\sim$0.6 pc). Then the visibility has been measured
to go down to $\sim$0.4 at a long baseline of 46 m at PA 45\degr\ with
VLTI/VINCI (Wittkowski et al. 2004). When combined together, a
multi-component model is favored for the K-band emission where a
significant part is coming from spatial radial scales clearly smaller
than $\sim$2.5 mas (HWHM; $\sim$0.16 pc; see Fig.2 of Wittkowski et
al. 2004).

\footnotetext{ Note that a elliptical Gaussian is probably an adequate
model for Type 2s rather than a ring, and the visibility curves for a
Gaussian and a ring are quite similar at the low spatial frequencies
before the first visibility minimum (zero) for the latter, if the
Gaussian HWHM is equal to the ring radius.}

These measurements are plotted in Fig.\ref{torus_pc} and
\ref{torus_mas}, where the estimation for the intrinsic luminosity
is taken from Pier et al. (1994).  Although there is inevitable
uncertainty in the luminosity estimation, the 2.5 mas upper limit
for the compact component from the VINCI measurement appears to fall
between $\rsub$ (for $\Tsub$=1500K and $a$=0.05 $\mu$m) and the
time-lag radius. The former is $\sim$4.3 mas for NGC 1068 ($\sim$0.27
pc), while the latter is $\sim$1.6 mas ($\sim$0.098 pc). The physical
interpretation for the compact component critically depends on the
correct innermost torus radius $\Rin$ to adopt. It might correspond to
e.g. the clumpiness of the torus if $\Rin$ is $\rsub$ (Wittkowski et
al. 2004; H\"onig et al. 2006). However, if $\Rin$ is the time-lag
radius, then the small-scale component ($<$2.5 mas) might nicely
correspond to the inner boundary region, extincted and partly seen
through a clumpy torus.

\section{Conclusions}

The high-spatial-resolution observations of Type 1 AGN tori in the
near-IR involves a contribution from the near-IR part of the big blue
bump emission, or the putative accretion disk emission.  We quantified
this contribution for the nuclear point sources in the HST/NICMOS
images of nearby Type 1 AGNs. At least for the available sample, the
K-band point-source flux appears to be dominated by a blackbody-like
emission from the innermost region of a torus, with only a small
contribution from the accretion disk.  For the sample studied, the
latter fraction is roughly $\sim$25\% or less.  We also used our
clumpy torus model to simulate the near-IR torus spectra, and conclude
that the estimation for the accretion disk component stays essentially
the same.

A theoretical prediction for the inner torus boundary size can be
given as a dust sublimation radius. We have shown that the time-lag
radius from the near-IR reverberation measurements is systematically
smaller by a factor of $\sim$3 than the predicted sublimation radius
with a reasonable assumption for graphite grains of sublimation
temperature 1500 K and size 0.05 $\mu$m in radius.  If the time-lag
radius is the correct innermost torus radius, this might indicate a
much higher sublimation temperature, but this appears to be disfavored
by the observed colors of the HST point sources studied here.  In this
case, the time-lag radius would suggest a dominance of much larger
grains in the innermost torus.  Alternatively the central engine
radiation might intrinsically be highly anisotropic, or there might be
a significant extinction in the equatorial plane between the central
engine and torus.

Based on the inferred dominance of the torus emission in K-band and
the expected innermost torus radius, we quantified the current
expectations for the near-IR interferometric observations of Type 1
nuclei.  These observations, with a long baseline of $\sim$100 m,
including the observations of NGC 4151 with different baselines,
should provide independent measurements for the innermost torus
radius. This will be important for the physical interpretation of the
current and future data obtained for AGN tori.


\begin{acknowledgements}

Authors would like to thank Robert Antonucci and Ari Laor for their
helpful comments and discussions.  This research used the facilities
of the Canadian Astronomy Data Centre operated by the National
Research Council of Canada with the support of the Canadian Space
Agency. This research has made use of the NASA/IPAC Extragalactic
Database (NED) which is operated by the Jet Propulsion Laboratory,
California Institute of Technology, under contract with the National
Aeronautics and Space Administration.  This research is partly based
on observations made with the NASA/ESA Hubble Space Telescope,
obtained from the Data Archive at the Space Telescope Science
Institute, which is operated by the Association of Universities for
Research in Astronomy, Inc., under NASA contract NAS 5-26555.

\end{acknowledgements}

\end{document}